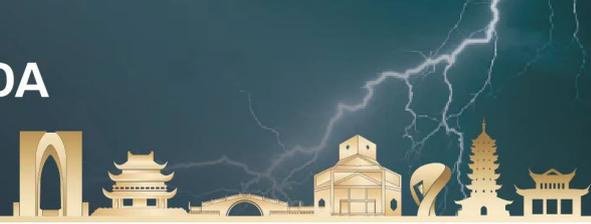

# Open-source FDTD solvers: The applicability of Elecode, gprMax and MEEP for simulations of lightning EM fields

Hannes Kohlmann
OVE Service GmbH
Vienna, Austria
h.kohlmann@ove.at
EPFL
Lausanne, Switzerland
hannes.kohlmann@epfl.ch

Dmitry Kuklin
NERC KSC RAS
Apatity, Murmansk Region, Russia
kuklindima@gmail.com

Wolfgang Schulz
OVE Service GmbH
Vienna, Austria
w.schulz@ove.at

Farhad Rachidi
EPFL
Lausanne, Switzerland
farhad.rachidi@epfl.ch

*Abstract*—In this study, the open-source finite-difference time-domain (FDTD) solvers gprMax, Elecode and MEEP are investigated for their suitability to compute lightning electromagnetic field propagation. Several simulations are performed to reproduce the results of typical field propagation scenarios that can be found in the literature. The results of the presented solvers are validated through comparison with reference field results corresponding to propagation over perfectly conducting and lossy ground. In most of the tested scenarios, all solvers reproduce the reference fields with satisfactory accuracy. However, close attention must be paid to the proper choice of the spatial discretization to avoid artificial numerical dispersion, and the application of the simulation cell boundaries, which can cause significant impairment of the results due to undesired reflections. Some cases of inaccurate FDTD results due to improper choices of parameters are demonstrated. Further, the features, the performance and limitations, and the advantages and drawbacks of the presented solvers are highlighted. For familiarization with the solvers' programmatical interfaces to initialize and run the simulations, the developed scripts are made available to the community in an openly accessible repository.

## I. Introduction

The computation of lightning electromagnetic (EM) fields has gained much importance within the last few decades. Numerical simulations allow to consider complex/realistic configurations which cannot be handled using (semi-)analytical approaches. Among the available numerical approaches, the FDTD method, derived from Maxwell's equations and first introduced by Yee [1], allows for a straightforward computation of the interaction of lightning EM fields with objects or structures such as mountains, towers, overhead power transmission lines, buried cables, railways, telecommunication infrastructure, etc. (see for example [2] and [3]). The FDTD method ([4]) yields full-wave time-domain results, which are directly comparable with real-life EM field measurements (e.g., obtained by E-field or H-field sensors used in lightning location systems, or measurement facilities of transmission system operators).

The paper sheds light on specific scenarios of lightning electromagnetic field propagation, where the properties of the EM fields (i.e., fast current rise times and long propagation distances) and the simulation domain are challenging for FDTD simulations. Another goal of this paper is to familiarize the reader with three open source FDTD software libraries, Elecode [1] [5], gprMax [2] [6], and MEEP [3] [7], which are all being actively developed and maintained, and at the same time to highlight their limits and some caveats specific to the individual frameworks, when applied to the analysis of lightning EM field propagation in general. To demonstrate artifacts that would occur when computational parameters, cell boundaries, etc., are not chosen appropriately, some of the parameters are intentionally specified beyond the permitted limits to provoke undesired simulation behavior and results.

## II. Methodology

### A. Generation of reference fields for validation

Before any qualitative interpretation of the results obtained by numerical computational methods, a preceding validation procedure to demonstrate the accuracy of their output is of high importance. Therefore, this step is performed to demonstrate, that the presented FDTD frameworks reproduce adequately expected results, which can be found in the literature, with satisfactory accuracy.

In terms of the impact of the ground parameters and structure on the resulting EM fields at different distances, the results of some basic, yet important, scenarios can be derived from fields that have propagated over perfectly electric conducting (PEC) ground. Subsequent filtering using specific formulations can account with very good accuracy for the impact of a lossy ground, ground stratification, etc., on the lightning EM fields above ground and underground ([8], [9], [10], [11]). From the relevant filtering functions, which are often described in (respectively derived from) the continuous frequency domain, the numerical representations have to be obtained for the discrete time domain using the inverse Fourier transform, which can be interpreted as the impulse response of the ground. Therefore, the fields over PEC can be computed in a first step and subsequently filtered using the time-domain convolution of the given impulse responses. Assuming that these obtained fields are good approximations of the true fields, they serve as the theoretical reference, to which the full-wave time-domain solutions yielded by the FDTD solvers can be compared. Ideally, the results should match. However due to limitations inherent to the FDTD method, and respectively the limitations specific to certain solvers, the deviations from the ideal solutions may be observed.

---

[1] https://gitlab.com/dmika/elecode
[2] https://github.com/gprmax/gprmax
[3] https://github.com/nanocomp/meep

The reference H- and E-fields over PEC ground were obtained numerically in a first step by using the method presented in [12], where the contributions of small current dipoles along the channel are summed up at an observation point. To represent the lightning return stroke channel, the modified transmission line model with exponential decay (MTLE, [13] and [14]) was used. The parameters of the model were set to $\lambda = 2$km (exponential decay of the RS current with height above the channel base), $v_{RS} = 1.5 \cdot 10^8$ m/s (RS wavefront speed). The considered channel base current is of Heidler-type,

$$I(0,t) = \frac{I_1}{\xi_1} \frac{\left(\frac{t}{\tau_{11}}\right)^{n_1}}{\left(\frac{t}{\tau_{11}}\right)^{n_1}+1} e^{-\frac{t}{\tau_{12}}} + \frac{I_2}{\xi_2} \frac{\left(\frac{t}{\tau_{21}}\right)^{n_2}}{\left(\frac{t}{\tau_{21}}\right)^{n_2}+1} e^{-\frac{t}{\tau_{22}}}, \xi_i = \exp\left(-\frac{\tau_{i1}}{\tau_{i2}} \left[n_i \frac{\tau_{i2}}{\tau_{i1}}\right]^{1/n_i}\right).$$

The parameters were chosen to form a channel base current that has the characteristics of a typical subsequent RS, with $I_1 = 10.7$ kA, $\tau_{11} = 0.25$ μs, $\tau_{12} = 2.5$ μs, $I_2 = 6.5$ kA, $\tau_{21} = 2$ μs, $\tau_{22} = 230$ μs and $n_1 = n_2 = 2$ (see [15]). The corresponding current waveform is depicted in Fig. 1.

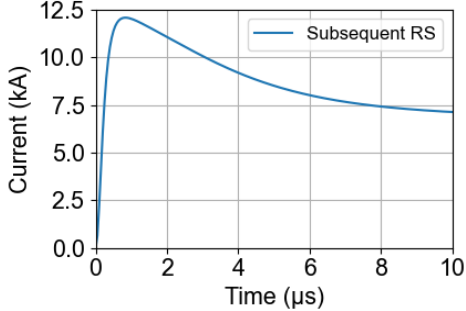

Fig. 1: Typical subsequent return stroke current waveform

### B. FDTD simulation setup

This section describes, how simulations are initialized using the three different solvers. For the sake of brevity, we only highlight the most important points, which have to be considered for a well-posed simulation initialization and can otherwise lead to confusion or a lengthy search for errors when the user is not familiar with the potential pitfalls. For the rest, the straight-forward initialization procedures can be found in the code examples, which are provided in the openly available repository:

https://gitlab.com/OVE_ALDIS/lightningfdtd

Common to all FDTD solvers is the proper choice of the discrete time step. It mu st fulfill the Courant-Friedrich-Lewy (CFL) convergence condition [16], which for three dimensions and uniform discretization ($\Delta x = \Delta y = \Delta z$) reads $\Delta t \leq \Delta x/c_0 \cdot 1/\sqrt{3}$, and thus imposes an upper bound (depending on the discrete spatial discretization $\Delta x$ of the FDTD grid), beyond which FDTD algorithms become unstable and the resulting fields diverge. In MEEP, the default value to guarantee convergence, and therefore stability, is $\Delta t = 0.5 \cdot \Delta x/c_0$, $c_0$ being the speed of light in vacuum, although it can be adjusted using the parameter "Courant". In gprMax and Elecode the time step $\Delta t$ is automatically computed according to the chosen spatial discretization, whereby an additional factor < 1 can be used to further reduce the time step to ensure stability. The spatial step $\Delta x$, in turn, must be chosen such that the wavelengths of the highest occurring frequencies are appropriately resolved in space. If they are insufficiently resolved, the numerical dispersion inherent to the FDTD method (see for example [4] for a detailed investigation) will lead to strong oscillations in the results, as will be demonstrated later in Section IV. A rule of thumb, which based on our experience leads to reasonable results, is to specify a discretization of about $1/10^{th}$ of the 10-90% rise time of a lightning EM wave front. For the considered current (see Fig. 1), the 10-90% rise time is about $\tau_R = 0.33$ μs, which corresponds to a spatial extent of $\lambda = \tau_R \cdot c_0 \approx 100$ m. Although not completely free of numerical dispersion causing oscillations, a spatial discretization of $\Delta = 10$m will yield good results. Since a lossy ground has the property of filtering higher frequencies, this criterion can be relaxed to up to $\Delta x = 50$ m when the ground conductivity is $\sigma = 1$ mS/m, which is essential to reduce the computational burden for 3-D FDTD simulations of large domains ranging over several hundred kilometers. In particular, the user is forced to use a coarse grid discretization for long propagation distances, since the random-access memory (RAM) is a limited resource. In three-dimensional FDTD simulations, increasing the spatial step by a factor of 2 results in $2^3 = 8$ times more nodes that must be stored and computed. Thus, the RAM demand increases by a factor of 8, and the computation time accordingly. On the other hand, a small discretization (high resolution) may be required if 3-D models shall be represented with sufficient detail, which may lead to the infeasibility of the FDTD method.

Another property that all FDTD algorithms have in common are the natural boundary conditions given by the edge of the cell, outside which the field values are necessarily zero. These boundaries behave like perfect electric conducting planes, completely reflecting the fields upon their incidence. However, it is usually desired to find formulations for the update equations at the cell boundaries, which, in the ideal case, should completely absorb the fields. These boundary layers are called 'perfectly matched layers' (PML). Various methods, some of which are also used in the presented FDTD solvers, are treated in [4]. All online documentations of the solvers contain sufficient information and references on their specific PML implementation. In this study, the performance of the solvers' default PMLs were tested for their capability to suppress reflections in terms of the selected thickness, respectively one set of parameters to specify the Elecode CPML formulation, which was empirically found to perform well. A deeper inspection of the particular PML implementations and possible parameter optimizations is beyond the scope of this work.

## III. SIMULATION RESULTS

### A. Validation of results above a flat ground

**PEC ground:** The simulated H- and E-field at a close-distance observation point of 1 km, are depicted in Fig. 2. The results of all three solvers are in perfect agreement with the reference fields (solid lines). It is particularly obvious from the E-field is that the FDTD method, due to its derivation from Maxwell's equations, can reproduce the effect of the electrostatic component, which is visible as the gradual increase of the E-field in the late-time response. The fields in Fig. 2 were computed using a spatial discretization of $\Delta x = 5$ m in Elecode and gprMax, and $\Delta x = 10$ m in MEEP[4].

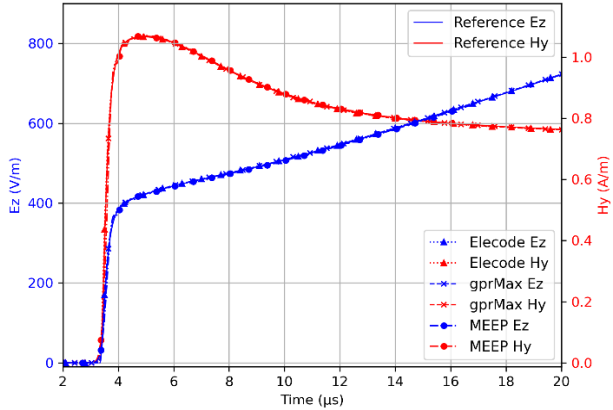

*Fig. 2: Fields at a distance of 1 km above a perfectly electric conducting ground. $\Delta x = 5$ m, $\Delta t = 8.67$ ns (Elecode and gprMax) and $\Delta x = 10$ m, $\Delta t = 8.67$ ns, $\Delta t = 17.33$ ns (MEEP).*

**Lossy ground:** The following computations were performed for flat and homogeneous lossy ground with conductivity $\sigma = 1$ mS/m and relative permittivity $\varepsilon_r = 10$. The validation was done for the horizontal electric field $E_x$ at 10 km distance, both at the ground surface level and at a depth of 10 m below the ground surface. The reference $E_x$-fields were obtained by performing the following steps: 1) Computation of the vertical

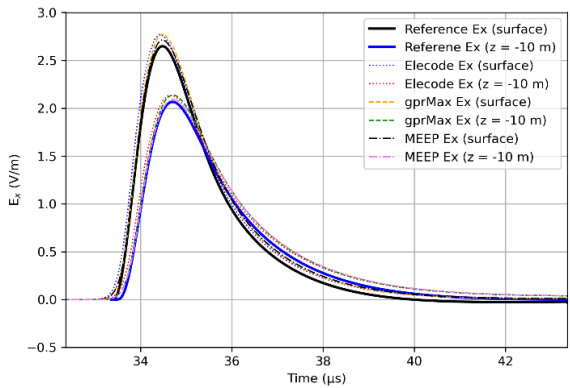

*Fig. 3: Horizontal electric field ($E_x$) at a distance of 10 km on the ground surface and at a depth of 10 m below the ground surface. The ground is flat, homogeneous, and lossy ($\sigma = 1$ mS/m, $\varepsilon = 10$). Spatial discretization $\Delta x = 10$ m, $\Delta t = 17.33$ ns.*

$E_z$-field above PEC ground, followed by 2) filtering of the $E_z$-field using formulations that account for the ground losses ([8] or [11]) and 3) using the resulting $E_z$-field above lossy ground to obtain the horizontal $E_x$-field at surface level through the wave tilt equation (see [9]). In step 4), the fields at a depth of 10 m below the ground surface were obtained by filtering the

$E_x$-field at the surface of lossy ground using Weyl's formulation (see [9]), which is a valid approximation for conductivities $\geq 1$ mS/m. The comparison of the FDTD results with the reference fields is depicted in Fig. 3. It is evident that the results obtained from the three different FDTD solvers are in good agreement with the reference fields, even though a coarse spatial discretization of 10 m was used, and therefore the underground sampling was performed just one discretization step below the ground surface.

*B. Perfectly matched layer performance*

A comparison of the results with respect to the choices of PML thickness obtained from the three solvers is shown in Fig. 4. In the simulations, the default PMLs of the solvers were used with thicknesses as shown in the legend. The best

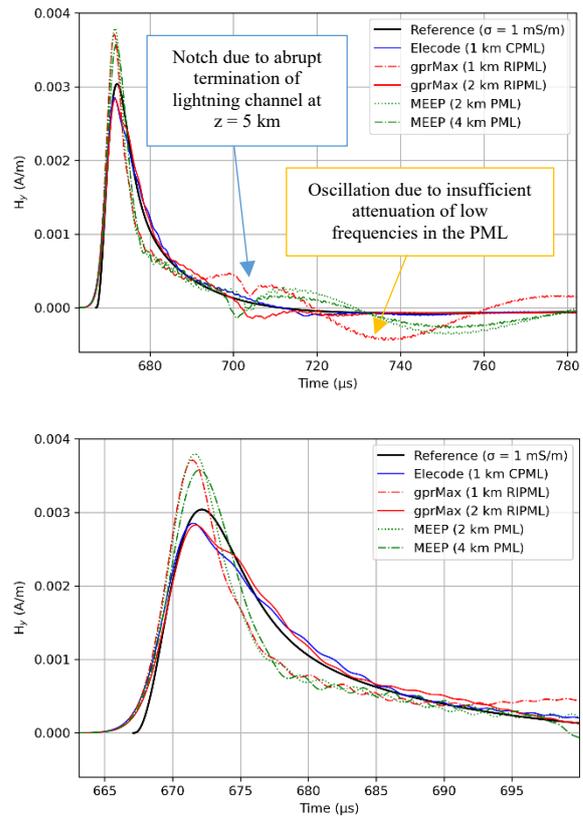

*Fig. 4: Comparison of the effectiveness of PMLs according to different thicknesses. $H_y$-fields at 200 km distance at the surface of lossy flat ground ($\sigma = 1$ mS/m, $\varepsilon_r = 10$). 3D-FDTD: $\Delta x = 50$ m, $\Delta t = 86.66$ ns. Extended view(top) and zoom view (bottom)*

results for simulations of 200 km field propagation, using a corridor of 18 km and a channel height of 5 km, are obtained when using the standard PML (CFS-PML) of at least 2 km thickness in gprMax, and CPML of 1 km in Elecode for a set of parameters which proved to perform well. In these cases, the field peak, the rise time and width are well preserved, while low-frequency fields are attenuated. The PML formulation used in MEEP produces detrimental reflections even with PMLs as thick as 4 km (changing the PML

---

[4] The coarser discretization of $\Delta x = 10$ m in MEEP is owed to the fact that the simulation was performed using MEEP with double precision floating points, which would not have permitted a discretization $\Delta x = 5$ m in terms of RAM demand.

conductivity profile from quadratic to linear or cubic did not lead to any improvement).

## IV. Discussion

Since the performed simulations were laid out to be validated through reference fields, the discussion will first focus on general remarks and caveats regarding the parameter choice and initialization, and the impact on the results, which the reader may encounter (Section IV-A). Secondly, a short discussion on the performance, limits, advantages and drawbacks of the three solvers will be given (Section IV-B).

### A. General

**Numerical dispersion:** If the spatial discretization is too coarse and the ground conductivity increases for example from $\sigma = 1$ mS/m to $\sigma = 3$ mS/m, numerical dispersion will become evident as undesired oscillations in the results. This is shown in Fig. 5, where a spatial discretization of $\Delta x = 50$ m was used. This phenomenon becomes even worse when even higher conductivities are used. Further, in Fig. 5 the impairment caused by the PML in MEEP (see Section is already observable at a distance of 100 km.

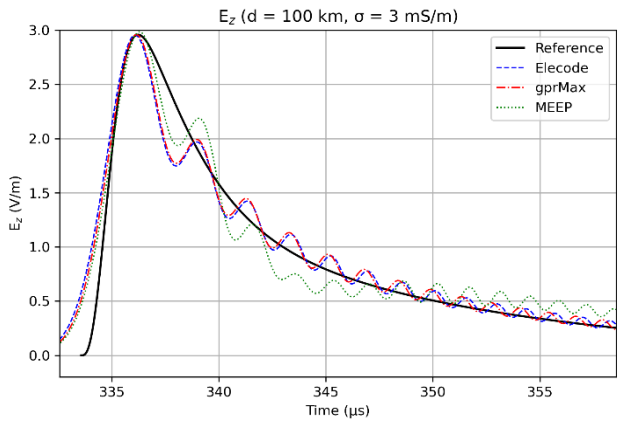

Fig. 5: *Numerical dispersion for fields at 100 km distance due to a higher conductivity value ($\sigma = 3$ mS/m, $\varepsilon_r = 10$) used with coarse discretization: $\Delta x = 50$ m, $\Delta t = 86.66$ ns. Sampled above flat, lossy ground).*

**PML:** Although not shown in this paper, the CPML employed in Elecode permits for corridors as narrow as 2 km in width given a certain parametrization. For gprMax the width had to be at least 5 km with 2 km wide PMLs using the standard PML without further specification. However, gprMax also supports parametrization of the CFS-PML and other, more elaborated, PML formulations (see Section IV-B, gprMax), where upon appropriate parametrization the absorption performance is expected to be significantly better. A test regarding the optimum parametrization, however, is beyond the scope of this study and is left for future investigation.

### B. Performance, limits, advantages and drawbacks of gprMax, Elecode, MEEP

**MEEP:** The initialization of simulations using MEEP, which only supports uniform (cubic) FDTD meshing, is straightforward thanks to the well-documented API. A detailed description of the MEEP API usage for lightning-related computations in the 2-D cylindrically symmetric domain was already given in [17]. The dimensionless quantities (conversion from MEEP units to SI units and vice versa) have to be handled at initialization and before any results can be compared to other results in SI units (see also [17] for details). Thus, any physical quantity like the conductivity must be converted from SI to MEEP units for the initialization, and sampled E- and H-fields back to SI units, which introduces a layer of complexity compared to the other frameworks under investigation. Further, by default, the software, which can be easily installed using the Anaconda Python developer tool, is built to operate with double precision floating points (FP). This doubles up the RAM demand, while for the considered simulation scenarios in the present study single precision would be sufficient. While building and installing the software with single precision FP is possible, it is not as simple as the installation of the default build. The material averaging functionality in MEEP leads to very good simulation results with relatively small stair-casing artifacts. In previous studies ([18], [19]), the implementation of, e.g., real 3-D terrain (although out of the scope of this work) has been done using discrete blocks or prisms to represent the elevation [5], which can take several hours depending on the extent of the domain. Another drawback of the framework is the perfectly matched layer (PML) implementation which causes strong reflections at oblique angles of incidence that are detrimental when considering long propagation distances. This problem is particularly exacerbated when narrow corridors are used. Also it is worth mentioning, that sampling of the $E_z$-field at an observation point in MEEP must be done at least one $\Delta x$ above ground, otherwise the fields are interpolated between the air-ground interface and thus distorted.

**gprMax:** Several advantages of gprMax are worth noting. The implementation permits to perform simulations using graphics processing units (GPUs, by utilization of the software layer CUDA), simply by adding a '-gpu' flag, once gprMax has been appropriately installed for GPU usage (for which a good instruction is provided in the online documentation). In particular, the gprMax simulations using a GPU enables up to 30 times faster simulations compared to the processor of typical consumer PCs (the comparison was made with an Intel® Core™ i7-9850H), specifically 14 times compared to the more performant CPUs like the Ryzen 9 3950x used in this study. Furthermore, having been developed for the scientific domain of ground penetration radar, the software permits for a simple initialization using dispersive materials (Debye, Lorentz, and Drude model), realistic soils, heterogeneous objects, and rough surfaces, which are all particularly interesting and useful for lightning electromagnetic scenarios as well. Arbitrary terrain scenarios can be initialized using 3-D index arrays (in HDF5 format) and an additional ASCII text-file that stores the material parameters corresponding to the indices in the HDF5 file. Aside from supporting rectilinear

---

[5] A simpler way to accomplish real 3-D terrain initialization was not found by the authors

FDTD meshing (with different spatial discretization Δx, Δy and Δz), sub-gridding will be supported as well in the future. This will allow to use locally finer discretization, if for example the field-terrain or field-object- interaction is of main interest in the simulations. The selectable PML formulations are based on a complex frequency shifted (CFS-) PML ([20]–[22]), which allow for very good performance both in terms of suppressing waves of low frequency, evanescent fields and obliquely incident fields, when the parametrization is done correctly. A proper choice of parameters is, however, not straight-forward. As described in the online documentation, the default parameters are α=0, κ=1, and σ is automatically determined by gprMax.

**Elecode:** The CPML formulation, which is implemented as described in [23], of the FDTD cell boundary enables to create very narrow corridors while retaining undisturbed fields, which is a strength of this framework. However, this PML is most difficult to handle in terms of the parameter choice. Furthermore, the main focus of this software has been initially targeted at grounding scenarios, considering frequency dependent (Debye-dispersive) soil and the implementation of arbitrarily oriented conductors (oblique thin wires representation, see [24]) embedded therein (see [25]). These conductors can be readily initialized with rather simple instructions in the ASCII-based initialization text file to represent transmission lines, buried cables and grounding conditions. Elecode only supports uniform (cubic) FDTD meshing.

**Simulation performance:** For the comparison of computation performances, a simulation of the domain size (218 x 20 x 7.6) km (= 265 MCells for Δx = 50 m) was performed using the three solvers on various CPU-/GPU-architectures. The memory demand (single precision floating point) of that simulation is about 19 GB. The results are summarized in terms of simulation time and time-steps per second. The used processor types / architectures were: Amazon Web Services (AWS) r6a instance (3rd generation AMD EPYC), AMD Ryzen 9 3950x (fastest 16-core CPU released in Nov. 2019) and an AWS g5 instance (NVIDIA A10G Tensor Core GPU with 24 GB of GDDR6 memory). The prices for the AWS cloud computing instances (running Linux OS) are 0.23$/hour for the r6a.xlarge instance with 32 GB RAM and 1.22$/hour for the g5.2xlarge instance with 32 GB RAM and 24 GB GPU memory (state end of March 2023).

V. CONCLUSION

In this study, a set of simulations for simple scenarios has been performed using three different openly available FDTD solvers, to demonstrate their applicability for lightning electromagnetic field computations. The results obtained by the FDTD solvers were validated against reference fields obtained using well-established formulations and approximations found in the literature. The goal was to highlight a few scenarios, for which the FDTD solvers yield accurate results, given the constraints that are either inherent to the application the FDTD method in general or more specific to the FDTD solvers under investigation. The simulation scripts were made openly accessible to the community to reproduce the results and to allow for further testing by the interested reader.

A caveat that generally applies while dealing with FDTD simulations is that, due to the multitude of options that FDTD frameworks offer and the peculiarities that prevail for the individual implementations of the FDTD algorithm, the results must be interpreted with utmost caution. Ideally, the plausibility of the results should be tested against some well-known reference fields as far as the analytical methods permit, before simulations of more complex domains are performed. In other words, the initialization of the domain with respect to its dimensions, the boundary conditions (type, thickness, etc.), the necessary spatial discretization depending on (a) the highest frequency components of the source functions, and/or (b) the properties of the terrain, the available RAM, etc., all play a role on whether a simulation can yield correct results or not.

Overall, the set of the presented openly available software frameworks, and the plurality of different relevant analysis tools that come along with them, offer a great opportunity to train scientists and enhance their (practical) understanding of the FDTD simulation method. The simulation scripts were made openly accessible in a Gitlab repository:

https://gitlab.com/OVE_ALDIS/lightningfdtd

| Performance / Processor | Seconds per iteration | Initialization time (HH:MM:SS) | FDTD solving time (HH:MM:SS) |
|---|---|---|---|
| SOLVER (N threads) → | Elecode gprMax MEEP | | |
| 3rd gen AMD EPYC | 1.55 sec/it (4) | 00:00:08 | 04:33:46 |
|  | 2.11 sec/it (2) | 00:01:06 | 06:05:34 |
|  | 1.75 sec/it (2) | 00:07:00 | 05:09:54 |
| AMD Ryzen 9 3950x | 1.53 sec/it (10) | 00:00:07 | 4:30:01 |
|  | 1.2 sec/it (10) | 00:01:00 | 03:27:09 |
|  | 1.62 sec/it (4) | 00:05:04 | 04:46:38 |
| A10G | X | X | X |
|  | 86.2 ms/it | 00:01:53 | 00:15:01 |
|  | X | X | X |

*Table 1: Performance comparison*